\documentclass[%
 reprint,
superscriptaddress,
 amsmath,amssymb,
 aps,
prb,
]{revtex4-1} %revtex4-1 or 4-2?

\usepackage[utf8]{inputenc}
\usepackage[english]{babel}
\usepackage{graphicx}% Include figure files
\usepackage{dcolumn}% Align table columns on decimal point
\usepackage{bm}% bold math
\usepackage{braket}
\begin{document}

%\preprint{APS/123-QED}

\title{Direct Spectroscopic Observation of Berry Phase Interference in the Ni$_4$ Single-Molecule Magnet}

\author{Brendan C.~Sheehan}%$^{1,2}$}%
\affiliation{Department of Physics and Astronomy, Amherst College, Amherst, MA 01002, USA}
\affiliation{Department of Physics, University of Massachusetts Amherst, Amherst, MA 01003, USA}%
\author{Robert Kwark}%$^{2}$}
\affiliation{Department of Physics and Astronomy, Amherst College, Amherst, MA 01002, USA}
\author{Charles A.~Collett}%$^{2}$}
\affiliation{Department of Physics and Astronomy, Amherst College, Amherst, MA 01002, USA}
\affiliation{Department of Physics, Muhlenberg College, Allentown, PA 18104, USA}
\author{Thomaz A. Costa}
\affiliation{Instituto de Qu\'imica, Universidade Federal do Rio de Janeiro, Rio de Janeiro, RJ 21941-909, Brazil}
\author{Rafael A.~All\~{a}o Cassaro}
\affiliation{Instituto de Qu\'imica, Universidade Federal do Rio de Janeiro, Rio de Janeiro, RJ 21941-909, Brazil}
\author{Jonathan R.~Friedman}%$^{1,2}$}
\affiliation{Department of Physics and Astronomy, Amherst College, Amherst, MA 01002, USA}
\affiliation{Department of Physics, University of Massachusetts Amherst, Amherst, MA 01003, USA}

\date{\today}

\begin{abstract}
Berry phase effects in spin systems lead to the suppression of tunneling effects when different tunneling paths interfere destructively.  Such effects have been seen in several single-molecule magnets (SMMs) through measurements of magnetization dynamics, where the experimental signal may arise from the contributions of numerous energy levels.  Here we present experimental measurements of Berry phase interference effects that are determined through electron-spin resonance on a four-fold symmetric SMM.  Specifically, we measure transitions between tunnel-split excited states in the Ni$_4$ SMM in the presence of a transverse field in the hard plane of the crystalline sample. By using a home-built rotation apparatus, the direction of the sample can be changed \textit{in situ} so that that the field direction can be swept through the entire hard plane of the sample.  When the field is in certain directions in the plane, we observe a splitting of the transition, a hallmark of Berry phase interference.  The experimental results are well reproduced by numerical simulations, and fitting of the data provides information about the effects of dipolar interactions and sample misalignment.
\end{abstract}

\maketitle

\section{Introduction}

Single-molecule magnets (SMMs) are zero-dimensional magnetic systems with $S >1/2$ with an energy barrier that separates spin states and leads to slow over-barrier relaxation at low temperatures.  A crystal of SMMs typically has ${\sim}10^{15}$ molecules, which are sufficiently separated due to the  presence of non-magnetic ligands so that intermolecular interactions are too weak to induce ordering and the crystal then behaves as an ensemble of spins.  These molecules exhibit many different kinds of quantum behavior, including quantum tunneling of magnetization.~\cite{friedmanSingleMoleculeNanomagnets2010} Furthermore, SMMs are attractive candidates for investigation as qubits due to their chemically tunable properties.

One particularly fascinating property of SMMs is Berry phase interference, a phenomenon in which multiple tunneling paths interfere coherently to enhance or suppress tunneling.  In 1993, Garg showed that tunneling can be ``quenched" by  destructive interference between the tunneling paths in spin systems with biaxial symmetry when the  field is applied along the spin's hard axis.\cite{gargTopologicallyQuenchedTunnel1993} The location of a quench in three-dimensional magnetic-field parameter space is known as a diabolical point.  Four-fold symmetric spin systems have also been shown to be capable of producing a similar effect.\cite{parkTopologicalQuenchingSpin2002,kimQuantumPhaseInterference2002,foss-feigGeometricphaseeffectTunnelsplittingOscillations2009}   Berry phase interference in SMMs was first observed experimentally by Wernsdorfer and Sessoli,\cite{wernsdorferQuantumPhaseInterference1999} who found an oscillating tunnel splitting when the applied field was aligned with the hard axis of the Fe$_8$ SMM, leading to a quenching of tunneling when the field produces destructive interference between paths. Since that observation, Berry phase interference in SMMs has been observed in a variety of systems,\cite{friedmanSingleMoleculeNanomagnets2010} including several variants of the Mn$_{12}$ SMM.\cite{wernsdorferQuantumPhaseInterference2002,lecrenQuantumTunnelingQuantum2005,adamsGeometricPhaseInterference122013} Futhermore, other flavors of SMMs have shown evidence of geometric phase interference, including half-integer-spin SMMs,\cite{wernsdorferQuantumPhaseInterference2005} trigonal SMMs,\cite{atkinsonThreeLeafQuantumInterference2014} an antiferromagnetic SMM,\cite{waldmannQuantumPhaseInterference2009} and other SMM-based systems including exchange-coupled SMM dimers.\cite{delbarcoQuantumSuperpositionHigh2004,leuenbergerBerryPhaseOscillationsKondo2006,wernsdorferInfluenceDzyaloshinskiiMoriyaExchange2008,quddusiAsymmetricBerryPhaseInterference2011} Many of these experiments based their observations on direct measurements of the magnetization of the molecule, and while clear interpretations of the results could be gleaned, the inference of spin dynamics from a thermodynamic quantity involving populations of multiple eigenstates can be challenging. Here we present a direct, spectroscopic observation of  geometric-phase interference using electron-spin resonance (ESR). Our work establishes unambiguous evidence for this form of interference in the Ni$_4$ SMM, a system with four-fold symmetry, and shows how the interference is modulated by the magnitude and direction of the transverse field within the hard plane of the molecule.

The Ni$_4$ SMM is composed of four Ni$^{2+}$ ions with $S=1$ ferromagnetically coupled to yield a total spin of $S=4$.\cite{yangExchangeBiasNi2003,lawrenceDisorderIntermolecularInteractions2008} The total number of states is therefore given by $2S+1=9$, ranging from $m=-4$ to $+4$.  Its effective ``giant-spin" Hamiltonian can be written as
\begin{equation}
\label{ham}
\mathcal{H} = -DS_z^2 - AS_z^4 + g_z\mu_B B_z S_z + \mathcal{H}',
\end{equation}
where $D$ and $A$ are positive axial anisotropy parameters, and $B_z$ is the $z$-component of the applied magnetic field $\mathbf{B}$. In the absence of $\mathcal{H}'$, the magnetic quantum numbers $m$ are eigenvalues of the Hamiltonian. At zero field, states $\ket{m}$ and $\ket{-m}$ are degenerate.  $\mathcal{H}'$ contains transverse terms that do not commute with $S_z$.  To a first approximation, the eigenstates can be described as $\ket{\pm}_m=\left(\ket{m}\pm\ket{-m}\right)/\sqrt{2}$, leading to tunneling between $m$ states.\cite{supp} The associated energies of these states can be labelled $E_{m,\pm}$. $\mathcal{H}'$ for Ni$_4$ is
\begin{equation}
\label{transham}
\mathcal{H}' = C(S_+^4 +S_-^4) + g\mu_B \mathbf{B}_\perp \cdot\mathbf{S},
\end{equation}
where $\mathbf{B}_\perp=B\sin\theta\left(\cos\phi\mathbf{\hat{x}} + \sin\phi\mathbf{\hat{y}}\right)$ is the transverse magnetic field.

\begin{figure}
    \centering
    \includegraphics[width = 0.48\textwidth]{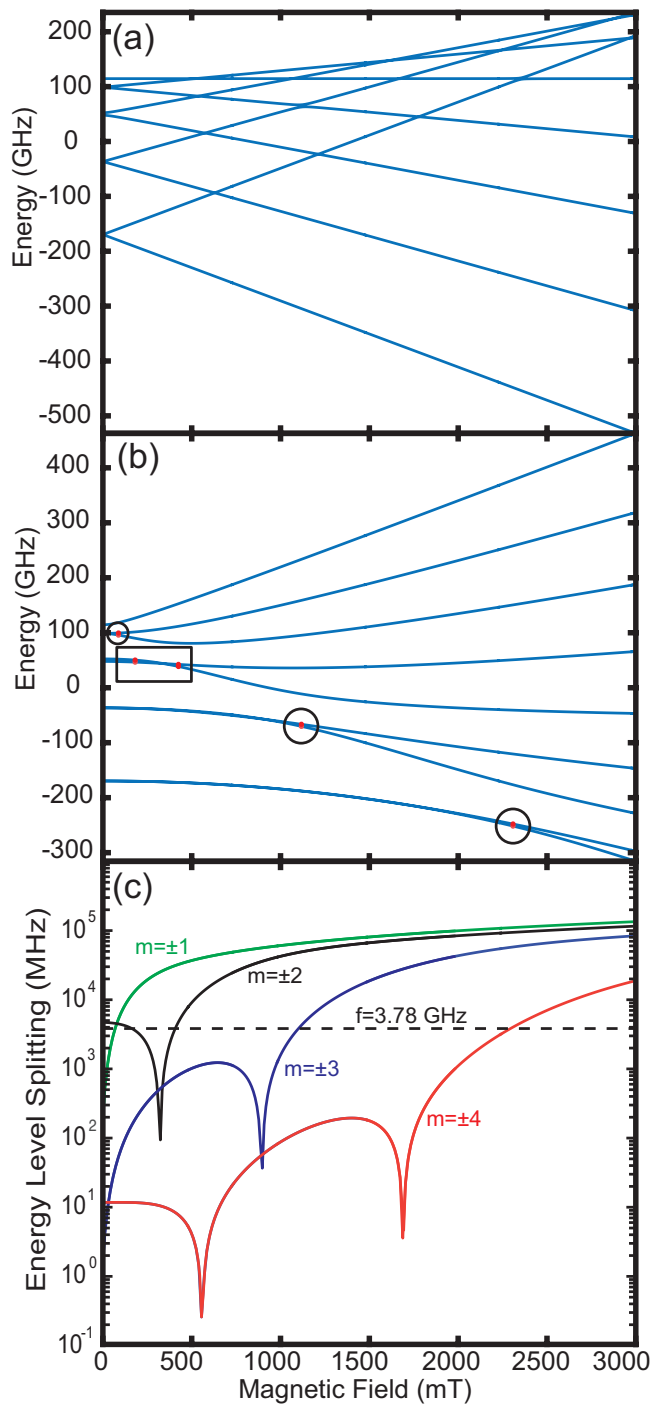}
    \caption{Field dependence of energy levels for the Ni$_4$ SMM.  (a) Level diagram for the field applied along the easy (z) axis. (b) Level diagram for a transverse field applied along a hard (x or y) axis. Observable transitions in perpendicular-mode ESR at a frequency of $f = 3.78$~GHz are indicated. The box shows the region explored in more detail in Fig.~\ref{levelsplot}. (c) The tunnel splitting (on a log scale) of the four pairs of levels as a function of transverse field along the hard axis.  The horizontal dashed line corresponds to a radiation frequency of 3.78~GHz.  Resonance occurs when this line intersects any of the tunnel splitting curves.}
    \label{full_levels_plot}
\end{figure}

It is important to note that different components of the magnetic field play distinct roles in SMMs.  A component along the z (easy) axis shifts the energies, as shown in Fig.~\ref{full_levels_plot}(a), producing a mostly linear dependence on field.  In contrast, components transverse to z contribute to $\mathcal{H}'$ and affect the tunneling.  Fig.~\ref{full_levels_plot}(b) shows the levels as a function of field applied along the x axis.  The splitting between pairs of nearly degenerate levels $\Delta_m=\left|E_{m,+}-E_{m,-}\right|$, known as the tunnel splitting, is shown in  Fig.~\ref{full_levels_plot}(c) as a function of transverse field, illustrating the effect of Berry phase interference of tunneling paths. Without the modulating effect of Berry phase interference on the tunnel splitting, an increasing $\mathbf{B}_\perp$ (Eq. \ref{transham}) suggests a monotonic increase in the tunnel splitting.

\begin{figure}[!htb]
\centering
\includegraphics[width = 0.48\textwidth]{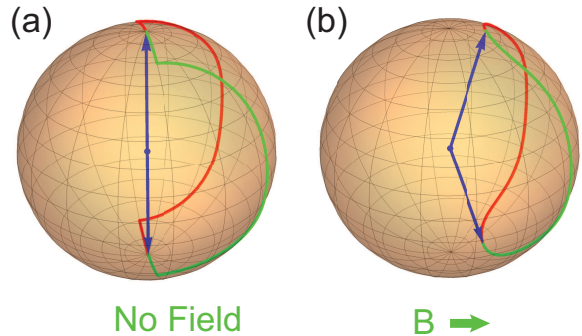}
\caption{(a) Bloch sphere showing spin-tunneling paths at zero external field. The blue arrows show spin directions corresponding to the ground state. The two paths, shown in red and green, correspond to a pair of least-action instanton paths. In zero-field, there are four degenerate instanton paths that interfere; only two are shown for clarity.  (b) Bloch sphere showing spin tunneling paths when an external field $B$ is applied parallel to a hard axis of the crystal. The applied alters the instanton paths and therefore the solid angle subtended by the paths. In each case, the solid angle is proportional to the Berry phase. Instanton solutions were calculated based on work published in Ref.~\onlinecite{foss-feigGeometricphaseeffectTunnelsplittingOscillations2009}.  For illustration purposes, highly exaggerated transverse anisotropy parameters were used in the calculations.}
\label{figberry}
\end{figure}

Berry phase interference occurs when there are multiple least-action (instanton) paths for tunneling between states.  As is generally true in quantum systems, the complex amplitudes of these paths must add, allowing for constructive and destructive interference between paths, depending on the relative phase associated with the paths.  In a spin system, the paths can be described as trajectories along the Bloch sphere connecting energy minima located at (or near) the poles. Ni$_4$ has four-fold rotational symmetry (\emph{cf.}~Eq.~\ref{transham}) and thus in zero field there are four least-action paths for tunneling. Two of these paths are shown in Fig.~\ref{figberry}(a); the other pair is hidden for clarity.  Symmetry ensures that each  path has the same amplitude, but they will have different geometric phases, giving rise to interference. The geometric phase is proportional to the solid angle between adjacent paths.  Application of a magnetic field would, in general, break the rotational symmetry of the system, suppressing the interference. However, if the field is applied along the x or y axes (the hard axes), a reflection symmetry is maintained so that for any tunneling path there is another with the same amplitude but different phase.  Fig.~\ref{figberry}(b) shows the paths when $\mathbf{B}_\perp$ is increased.  The solid angle subtended by the paths, on the right-hand side of the Bloch sphere, becomes smaller.  Because the geometric phase is proportional to this solid angle, an increasing transverse field causes the solid angle to decrease and so the interference is modulated between constructive and destructive.  When the interference is completely destructive, the tunneling is suppressed, leading to the sharp dips in the tunnel splittings shown in Fig.~\ref{full_levels_plot}(c).  This is the primary signature of Berry phase interference:  the tunnel splitting oscillates as a function of the transverse field instead of monotonically increasing.

\begin{figure}[!htb]
\centering
\includegraphics[width = 0.48\textwidth]{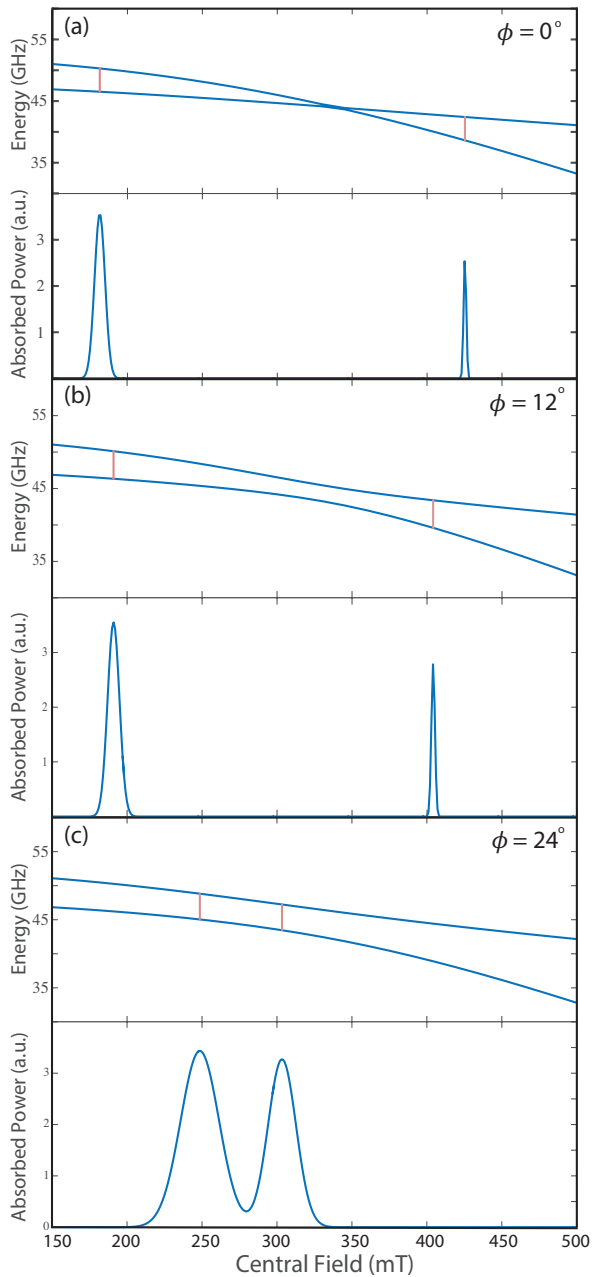}
\caption{Levels and simulated spectra for field in the hard plane, focusing on the boxed region of Fig.~\ref{full_levels_plot}. (a) the field lies along a hard axis, allowing  complete destructive interference between tunneling paths; tunneling is quenched, resulting in the degeneracy between levels at $B\approx340$~mT. Transitions at fields above and below this quench can be observed, as shown in the simulated spectrum.  As the field is rotated within the hard (X-Y) plane (b and c), the interference is suppressed, the degeneracy is lifted and the transition peaks move together. In each panel, the angle $\phi$ represents the angle between the field and the (hard) x axis.}
\label{levelsplot}
\end{figure}

Berry phase interference is reflected in the transverse-field dependence of the energy levels shown in Fig.~\ref{full_levels_plot}(b).  As the transverse field increases, the tunnel splitting varies and resonance with the applied radiation field will take place when $h f=\Delta_m$, where $f$ is the applied RF frequency.  Such transitions involve radiative coupling of the $\ket{+}_m$ and $\ket{-}_m$ states.  These states only have a matrix element for the $S_z$ component of spin. To wit, $_m\bra{+}S_z\ket{-}_m=m$, to a first approximation.\cite{supp}   An exact calculation of the eigenstates and matrix elements leads to substantially similar conclusions.\cite{supp}  The substantial $S_z$ matrix element implies that the radiation magnetic field should lie parallel to the easy axis of the sample.  With such an experimental configuration, the tunnel splitting can then be directly probed by ESR.  Transitions for $m=1,2,3,4$ are observable, indicated by the small red arrows in Fig.~\ref{full_levels_plot}(b).  However, in our experiments, the Berry phase oscillations for $m=3,4$ occur on energy scales too low to be observed and the tunnel splitting can be measured only at fields larger than the last quench.  The $m=1$ transition has a single quench at zero field, giving it the character of a Zeeman doublet.  In contrast, the $m=2$ transition has a clearly observable non-trivial field dependence.  With the field along the hard axis, the tunnel splitting goes to zero at $B\approx340~mT$, the consequence of complete destructive interference of tunneling paths.  This can be seen more clearly in the upper panel of Fig.~\ref{levelsplot}(a), which shows a zoomed-in view of the boxed ($m=2$) region in Fig.~\ref{full_levels_plot}(b).  The level degeneracy (diabolical point) gives rise to a clear ESR signature: two transitions can be observed for the same pair of levels, one at a field below the quench and one above.  The lower panel in Fig.~\ref{levelsplot}(a) shows a simulated spectrum with these transitions.  An important feature of Berry phase interference is that a field applied along a hard axis preserves the symmetry of the system.  By moving the transverse field (by an angle $\phi$, as defined in $\mathbf{B}_\perp$) away from the hard axis within the hard plane, the symmetry is broken and one path is favored over others, suppressing interference.  This is illustrated in Fig.~\ref{levelsplot}(b) and \ref{levelsplot}(c) for two values of $\phi$.  As $\phi$ increases, the degeneracy is lifted, giving rise to an avoided crossing.  The transitions move closer together and, for large enough $\phi$, eventually merge into a single transition.  Since the system has four-fold symmetry, the spectral dependence on $\phi$ should be periodic with period $\pi/2$.  Thus, by varying the magnitude and direction of the field within the hard plane, the unique spectral features of the Berry phase interference can be mapped out.

In this work, we spectroscopically measure the tunnel splittings in Ni$_4$, in particular the $m=2$ transition, and follow the behavior of the observed ESR spectra as the direction of the field is varied in the hard plane.  We qualitatively and quantitatively observe the expected signatures of Berry phase interference discussed above, providing strong evidence of this effect in the  Ni$_4$ SMM.

\section{Experimental Methods}

We directly observed the transitions in Ni$_4$ through low-temperature continuous-wave (cw) ESR measurements. We developed a method of \textit{in situ} sample rotation to allow for consistent sample realignment between ESR spectrum measurements. We performed the ESR measurements within a Quantum Design Physical Property Measurement System (PPMS) cryostat, which contains a nine Tesla superconducting electromagnet.

\begin{figure}[!htb]
\centering
\includegraphics[width = 0.48\textwidth]{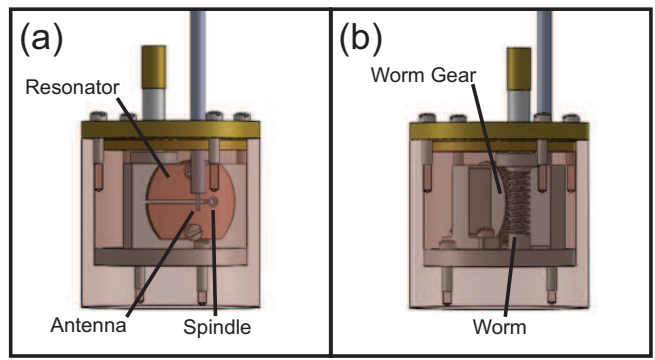}
\caption{CAD drawings of the rotator apparatus and resonator.  (a) View of resonator side of the apparatus, showing loop-gap resonator, antenna, and the end of the spindle.  The resonator is mounted using nylon screws. (b) View of the gear drive mechanism, showing worm and worm gear. As the worm gear is rotated, the sample located at the end of the spindle turns inside the loop of the resonator.}
\label{howto}
\end{figure}

The apparatus is designed to rotate a crystal of Ni$_4$ about its easy axis with the applied DC field in the hard plane and the RF field along the easy axis.  Figure \ref{howto} shows a CAD drawing of the heart of the apparatus, which sits within the sample chamber of the PPMS.  ESR spectra were obtained using a loop-gap resonator (LGR), which produces a uniform, strong RF magnetic field within the loop and has a resonant frequency of $\sim$3.78~GHz and a quality factor of $Q\sim1200$. ESR was performed in reflection mode with a single coaxial cable that runs the length of the sample chamber providing the source radiation and the reflected signal. Radiation coupling between coax and LGR was achieved through an antenna comprising an exposed section of the coax's inner conductor that is brought close to the gap of the LGR. Measurements of reflected power were obtained with a Keysight E5063A Vector Network Analyzer; data on the reflected power at resonance, the resonant frequency, and the quality factor were obtained as a function of magnetic field.\cite{supp}

\textit{In situ} rotation of the sample was achieved through a custom-designed, 3D-printed worm drive mechanism.  A stepper motor outside the cryostat turns a G10 rod that runs the length of the cryostat and attaches to the mechanism.  The rod turns the worm, which rotates the worm gear.  A spindle located on the axis of the worm gear extends into the loop of the LGR.  The Ni$_4$ crystal, roughly 1~mm in length, was placed on the end of the spindle and held in place with a small amount of vacuum grease.  The sample, which has a bipyrimidal shape, was carefully oriented to align the easy axis (long axis of crystal)  with the spindle's axis.  The LGR, sample and worm drive mechanism are contained inside a copper shield to prevent radiation losses that degrade the resonator $Q$.  Using this apparatus, all data from a single sample could be collected during a single cooldown.

Ni$_4$ was synthesized according to published procedures.\cite{yangFastMagnetizationTunneling2006} Importantly, Ni$_4$ contains two distinct conformational states (isomers) at low temperatures arising from distinct ligand geometries, which occur in roughly even proportions in the bulk crystal structure.\cite{collettPrecisionESRMeasurements2016}  This results in a doubling or broadening of the ESR spectral peaks. 

\begin{figure}[!htb]
\centering
\includegraphics[width = 0.48\textwidth]{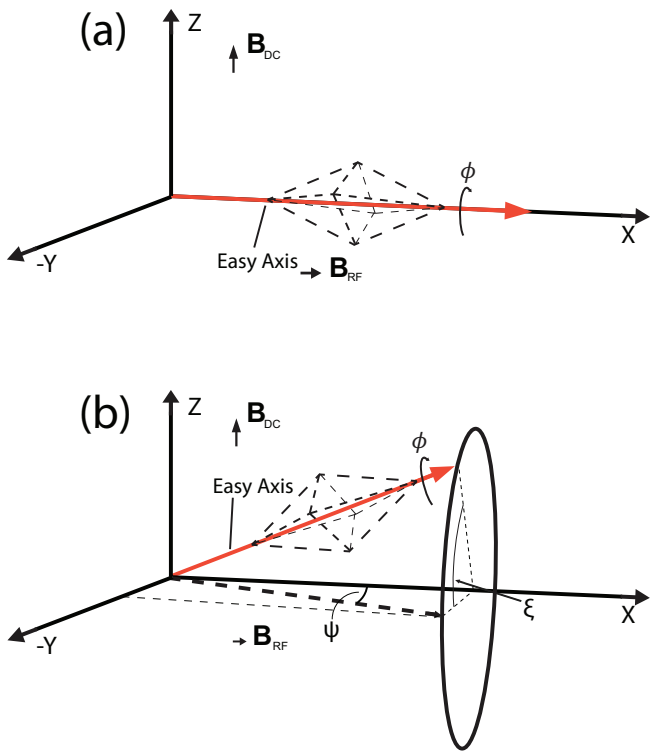}
\caption{(a) Ideal alignment of the crystal would have the easy axis along the X axis, and the crystal would rotate about its easy axis. In this case, the angle $\phi$ is equivalent the experimental rotation angle $\xi$. (b) Misalignment of the easy axis from the  X axis by $\psi$ results in the actual experimental rotation angle, $\xi$, being distinct from $\phi$.}
\label{figxi}
\end{figure}

To describe the experiment precisely, we use lower-case labels (x,y,z) to refer to the crystal axes and upper case (X,Y,Z) for the laboratory axes.  The DC field lies along the Z axis, and the RF field and spindle are parallel to the X axis.  Rotation of the spindle is characterized by an angle $\xi$. For a perfectly aligned sample in which the sample's z axis coincides with the X axis, $\xi$ is identical (up to a constant offset) to $\phi$, the angle between the (hard) x axis of the sample and the applied DC field (the Z axis), as shown in Fig.~\ref{figxi}(a).  In practice, however, there is a small misalignment $\psi$ of the easy (z) axis of the sample from the spindle axis, meaning that $\xi$ is not equivalent to $\phi$, as shown in Fig.~\ref{figxi}(b). When rotating the crystal by $\xi$, the sample still rotates by $\phi\approx\xi$, but the easy axis also wobbles from slightly above to slightly below the X-Y plane, meaning that the DC field has a small $\xi$-dependent component along the easy (z) axis (the effect of which is discussed below). For a given misalignment $\psi$, it is straightforward to use standard rotation matrices to describe the orientation of the crystal as a function of $\xi$.  The crystal orientation can be fully described in terms of $\psi$, $\xi$, and $\phi_0$, the value of $\phi$ at $\xi=0$. 

\section{Results \& Discussion}

\begin{figure}[!htb]
\centering
\includegraphics[width = 0.48\textwidth]{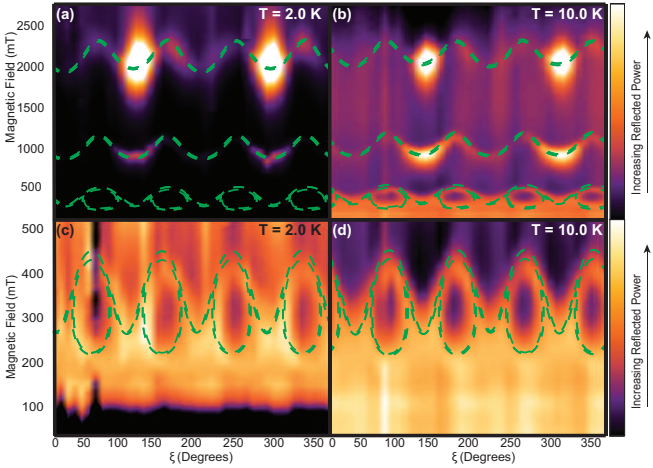}
\caption{Measured spectra, taken at 2.0~Kelvin (left side) and 10.0~Kelvin (right side). Each transition can clearly be seen to oscillate with changing $\xi$, and the intensity of the peak transitions oscillates, as well. (a) and (b) show the full-spectrum color plots at each temperature, as indicated; each shows the $m=3$ transition (which occurs near $\sim1000$~mT) and the $m=4$ transition (which occurs near $\sim2000$~mT). Green dashed lines are theory curves of the resonance-peak values for each transition. (c) and (d) show zoomed views of the low-field region of the spectra to highlight the interference effects observed as a function of $\xi$ in the $m=2$ transition, seen between 200~mT and 450~mT. The $m=1$ transition is also visible at $\sim$100~mT.  In each panel, the two conformational states of Ni$_4$ are calculated separately for the $m=2$ transition. %Note that in all panels, the quality factor $Q$ is scaled to the maximum value of the figure.
}
\label{colorplotsdata}
\end{figure}

We measured cw ESR field spectra at intervals in $\xi$ of 4.5$^\circ$, resulting in 81 spectra being taken over a full rotation.  The resonant frequency of the LGR was 3.78~GHz. Figure \ref{colorplotsdata} shows the change in reflected power $\Delta P$ as a function of field over a full range in $\xi$, at both T = 2.0 K (panels (a) and (c)) and T = 10.0 K (panels (b) and (d)).  Near 2000~mT, transition peaks represent the ground state tunnel-split transition ($m=4$), while the transitions near 1000~mT are the $m=3$ transition.  The $m=3$ and $m=4$ transitions each have two expected peaks for the two comformational states of Ni$_4$, but they lie very close together and cannot be clearly resolved.  At low fields, visible in panels (a) and (b), as well as in further detail in panels (c) and (d), are the peaks of the $m=2$ transition (the $m=1$ transition also appears faintly at ${\sim}100$~mT). The lower panels use a different color scale for clarity.  For certain values of $\xi$, there are two clear peaks visible in the 200 -- 400~mT range, corresponding to the bifurcation of the $m=2$ transition.  As $\xi$ is rotated through a full circle, the fourfold nature of the bifurcation appears clearly:  As $\xi$ is varied, the two peaks become closer together and eventually merge into one, behaving qualitatively as expected (\textit{cf.}~Fig.~\ref{levelsplot}).

In Fig.~\ref{colorplotsdata}, the highly visible $m=3$ (${\sim}1000$~mT) and $m=4$ (${\sim}2000$~mT) transitions reveal two obvious patterns.  The peak positions of both transitions shift as $\xi$ is varied, as expected when the transverse field sweeps through the hard plane of Ni$_4$.  In addition, the amplitude of the transitions changes dramatically.  This is due to the crystal misalignment (Fig.~\ref{figxi}(b)).  When the crystal easy (z) axis is in the X-Y plane, the DC field lies in the hard plane and the pairs of states can be well approximated by the superposition states $\ket{\pm}_m$, with a large radiative coupling within each pair.  However, as the sample is rotated, the easy axis leaves the X-Y plane, resulting in a small component of the DC field along the easy axis, $B_z$, that tends to localize the eigenstates, reducing the transition matrix elements between the two states so the transitions are suppressed.\cite{supp}  This effect is less pronounced at low fields since a small DC field results in a correspondingly small $B_z$ when the sample is rotated out of the X-Y plane.  

Superimposed on the data in Fig.~\ref{colorplotsdata} are curves corresponding to the theoretically predicted positions of the transitions.  Due to the closeness of the peaks of each conformational state in the Ni$_4$ SMM, the theoretical curves for the $m=3$ and $m=4$ transitions show the average predicted position of the peak positions.  In contrast, for the $m=2$ transition, theory predicts rather different results from the two conformational states and so the curves for both conformation states are presented.  Fig.~\ref{colorplotsdata} illustrates our fundamental finding: the periodic bifurcation of the $m=2$ transition (panels (c,d)), as predicted (Fig.~\ref{levelsplot}) due to the Berry phase interference and the existence of a diabolical point in this pair of levels.  The fact that the observed positions of the resonances agree with the theoretical predictions lends strong credence to this interpretation. The figure also demonstrates the four-fold symmetry of the peak positions, especially in the lower panels, with the pattern repeating every 90$^\circ$.  Also apparent, especially in the upper panels, is the effect of misalignment, which produces much stronger signals at two values of $\xi$ 180$^\circ$ apart, corresponding to the orientations when the sample's easy axis lies in the X-Y plane and $H_z\approx0$.  Since the $m=2$ transition corresponds to a transition between high-lying levels (cf.~Fig.~\ref{full_levels_plot}), these transitions become stronger at higher temperatures, as shown in the right panels of Fig.~\ref{colorplotsdata}.  Similarly, the $m=3$ transition is stronger at higher temperature while the $m=4$ transition, a ground-state transition, becomes weaker as the temperature is increased.  

\begin{figure}[!htb]
\centering
\includegraphics[width = 0.48\textwidth]{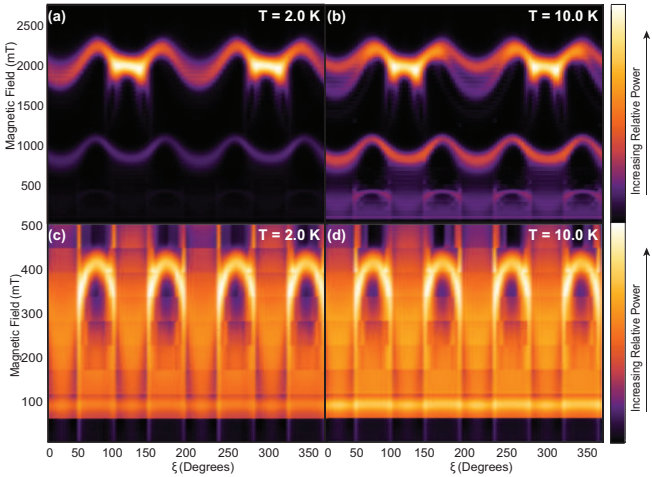}
\caption{Simulated best-fit spectra. %at best-fit temperatures of T=3.90~K (a) and (c) and T=12.53~K (b) and (d). 
As in Fig.~\ref{colorplotsdata}, (a) and (b) show simulations of the full spectra, while (c) and (d) show a close-up look at low fields to focus on the behavior of the $m=2$ transition.}
\label{colorplotssim} 
\end{figure}

\begin{figure}[!htb]
\centering
\includegraphics[width = 0.48\textwidth]{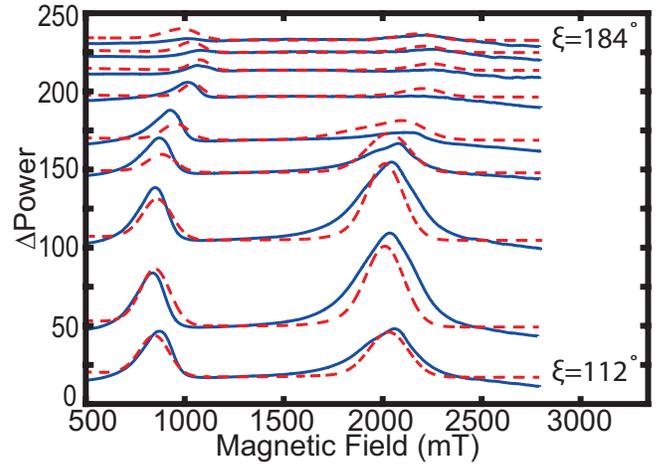}
\caption{Selected individual measured spectra, in this case at 10.0 Kelvin, as a function of $\xi$, shown as $\Delta P=P-P_{\text{back}}$, overlaid with corresponding simulated spectra as a function of applied field. $P_{\text{back}}$ is defined as the $P$ value of the background, measured at 220 mT for each spectrum. Solid blue lines represent data, while dashed red lines were generated using a best-fit simulation.  Low fields are ignored during the fitting process due to relatively low populations as compared to the lower energy $m=3$ and $m=4$ transitions.  The amplitude of the peaks in the data varies more than in the fits, but the peak positions are well-reproduced by the simulations. Note that in this figure, the spectra represent intervals of 9$^\circ$ in $\xi$, with the bottom and top curves labeled. The spectra are vertically offset for clarity.}
\label{selspec}
\end{figure}

\subsection{Simulations and Fitting}

Simulated spectra were produced using the EasySpin package for Matlab.\cite{stoll_easyspin_2006}  We used simulated spectra to fit our data using a least-squares method and extract relevant parameters (discussed below).  To reasonably reproduce our experimental spectra, we needed to account for various factors, including misalignment and the effects of dipole fields within the sample.  These factors have correlated effects, requiring them to be treated carefully.

To start, we take into account that the field $\vec{B}$ seen by a typical spin in the sample differs from the applied field $\vec{H}$: \begin{equation}
\label{Hialpha}
H_i=B_i-\alpha M_i(B_i,T),
\end{equation}
for the $i$th Cartesian component, where $\alpha$ characterizes lattice and demagnetization effects%with a value for Ni$_4$  that has been determined previously.
.\cite{chenObservationTunnelingassistedHighly2016}  The magnetization $\vec{M}$ has both a mean value $\vec{M}_0$ and a small random portion $\delta\vec{M}$:
\begin{equation}
\label{Mtot}
\vec{M}(\vec{B},T) = \vec{M}_0(\vec{B},T) + \delta\vec{M}.
\end{equation}

The former can be determined using basic statistical mechanical techniques:
\begin{equation}
M_{0,i}=\frac{g_i\mu_B\sum_{j=1}^9\braket{E_j|\hat{S}_i|E_j} e^{-E_j/k_BT}}{v \sum_{j=1}^9e^{-E_j/k_BT, }},
\label{mag}
\end{equation}
where $v$ is the unit cell volume of Ni$_4$.  The matrix elements and energies are found by diagonalizing the spin Hamiltonian (Eq.~\ref{ham}) for each conformational state to calculate the magnetization for that state.  The net magnetization is obtained by taking the average of the magnetizations of the two conformation states.  Hamiltonian parameters used for the calculations are based on previously determined values.\cite{chenObservationTunnelingassistedHighly2016}  

The random field $\delta\vec{M}$ is due to fluctuations of $M$ about equilibrium and arises from the configuration of neighboring spins.  Since the molecule is an easy-axis system, we assume the primary direction for the random dipole fields is the z axis and neglect the other components: $\delta\vec{M}=\delta M \hat{\mathbf{z}}$.  $\delta M$ is assumed to have a Gaussian-weighted distribution of width $\sigma$:
\begin{equation}
    \label{gauss}
    P\left(\delta M\right)=\frac{1}{\sigma\sqrt{2\pi}}e^{-\delta M^2/2\sigma^2}.
\end{equation}

Calculation of a spectrum proceeds as follows.  For a given applied field $H$, and orientation of the crystal (specified by angles $\psi$, $\xi$, and $\phi_0$), the components $H_i$ along the crystal axes are calculated.  Given a temperature $T$, for each value of $\delta M$ in a distribution of width $\sigma$, we numerically invert Eq.~\ref{Hialpha}, making use of Eq.~\ref{Mtot}, to determine the components of $\vec{B}$, the field experienced by a spin.  Using a range of values $\delta M$, the total (weighted) spectral response is then calculated for this field and the procedure is iterated over the full range of $H$ to obtain a complete spectrum.  In addition, the effects of $g$ strain can be included in the calculated spectrum.  These effects are essentially indistinguishable from those of crystal mosaicicity, as discussed below.  Iteration of this procedure over every value of $\xi$ creates a simulation of the full experiment.

\subsection{Fitting Results}

We implemented a least-squares fitting routine that considered the full behavior of the spectrum at every measured value of $\xi$. Fitting includes only the $m=3$ and $m=4$ transitions since the other observed transitions are too small to have a significant effect on the fits.  The results of fitting are shown in Fig.~\ref{colorplotssim}, and partially in Fig.~\ref{selspec}.  The fits reasonably reproduce the experimental results, although the amplitude of the peaks vary as a function of $\xi$ more in the actual data than in the fitted spectra, as shown in Fig.~\ref{selspec}.  While we do not have a definitive explanation for this discrepancy, we conjecture that it may result from sample heating by the radiation, as discussed further below. Nevertheless, the peak positions agree very well.  Remarkably, although the low-field data was not included in the fitting routine, the simulated spectra reproduce the observed interference effects in the $m=2$ transitions extremely well (cf.~Fig.~\ref{colorplotsdata} and Fig.~\ref{colorplotssim}(c,d)). This provides strong confirmation that we are observing the anticipated Berry phase interference in Ni$_4$.

Fitting parameters include the misalignment angle $\psi$; initial orientations of the crystal and apparatus, $\phi_0$ and $\xi_0$ respectively; the dipole Gaussian width $\sigma$; the magnetization factor $\alpha$; $g$ strain; and the temperature; as well as an overall scaling factor. We fit data taken at both  2~K and 10~K. Since the same sample was measured at both temperatures in a single cool down,  the difference $\delta =\phi_0 - \xi_0$ is the same for  both sets of data. We treat $\delta$ as the free parameter that, with the value of $\xi_0$ for each temperature, determines the value of $\phi_0$ for that temperature. Best-fit values of these parameters are given in Table~\ref{table}.  Parameters $\sigma$, $\alpha$ and $g$ strain represent intrinsic properties of the sample while $T$ is an essential property of the experiment.  In contrast,  $\xi_0$, $\delta$ and $\psi$ are ``accidental" properties relating the sample or apparatus alignment.  For completeness, the table also includes a value for a mosaic distribution width $\sigma_m$ that produces spectra (and therefore a fit) that is essentially identical to that obtained with the given value of $g$ strain.  We treat the mosaicicity as a Gaussian distribution in angular orientations of individual molecules within the crystal, where the center of the Gaussian represents the overall orientation of the crystal itself.  (Use of $g$ strain for fitting is computationally more efficient.)

The value of $\sigma$ obtained from the fitting is on the order of the nearest-neighbor dipole field for molecules in the crystal.  The fitted value of $\alpha$ agrees with the value determined in previous experiments on Ni$_4$.\cite{chenObservationTunnelingassistedHighly2016} While a $g$ strain on the order of 12\% is surprisingly large, it translates into a mosaic spread of $\sim0.3^\circ$ that is reasonable for molecular crystals.

\begin{table}[h!]
\centering
\begin{tabular}{||c|c||} 
 \hline
 \textbf{Parameter} & \textbf{Value} \\ 
 \hline\hline
 $\psi$ & $1.0(1)^\circ$\\%0.95^o\pm0.14^o$ \\ 
 \hline
 $\sigma$ & 14(3)~mT \\ 
 \hline
 $\alpha$ & 4(1) \\ 
 \hline
 T (2.0~K data) & 3.9(6)~K \\ 
 \hline
 T (10.0~K data) & 12.7(6)~K \\ 
 \hline
 $\xi_o$ (2.0~K data) & $106(5)^\circ$ \\ 
 \hline
 $\xi_o$ (10.0~K data) & $129(4)^\circ$ \\
 \hline
 $\delta$&$23(4)^\circ$\\
 \hline
 $g$ strain & 12(4)\% \\
 \hline
 $\sigma_m$ & $0.3(1)^\circ$ \\
 \hline
\end{tabular}
\caption{Fitting results for all free parameters. For each parameter, the value applies to both sets of data unless specified.}
\label{table}
\end{table}

The most significant deviations of fitting parameters from experiment are in the fitted temperatures, as mentioned above.  For the 2.0~K data, the fit temperature is 3.9~K, nearly twice the experimental temperature.  This may indicate an issue of sample heating by the applied microwave radiation.  Indeed, these experiments were done at high power (0~dBm) to obtain a good signal-to-noise ratio.  Heating by absorption of radiation (and emission of phonons) drives the system out of thermal equilibrium\cite{balPhotoninducedMagnetizationReversal2004,balNonequilibriumMagnetizationDynamics2005} and depends on the transition: more heating is expected for the ground-state $m=4$ transition.  Thus, use of a ``temperature" for a spectrum (or set of spectra) is heuristic and does not fully characterize the level populations of the system as a function of field.  In keeping with this interpretation, we find that for the 10.0-K data, the fit temperature of 12.7~K is a significantly smaller relative deviation, as one might expect: when the temperature is higher, the sample has a higher specific heat and better effective thermal coupling to the cryostat reservoir.  The amount of heating may further depend on $\xi$ because of changes in the matrix element as the sample's easy axis is rotated in and out of the X-Y plane, resulting in temperature changes that cannot be accounted for with a single value of temperature for a full set of data.  This may account for the disagreement between experimental and simulated peak amplitudes seen in Fig.~\ref{selspec}. 

\bigskip 
\section{Conclusion}

In this paper we have provided compelling evidence of Berry phase interference effects in the single-molecule magnet Ni$_4$. In particular, we explored an excited-state tunneling transition that shows a bifurcation at a given frequency as the applied magnetic field is swept. This doubling is dependent on the angle of the applied field relative to the hard axes of the crystal, and as such, modulation of these transitions occurred as the crystal was rotated. An \textit{in situ} method of sample rotation allowed investigation of the behavior as the field direction is swept through the hard plane of the sample, and showed the expected periodic bifurcation of the resonances, the hallmark of Berry phase interference. 

Furthermore, simulations of the ESR spectra clearly reproduce the bifurcation effects and show agreement with data from other excited transitions. We found that sample misalignment and the effects of dipolar interactions between molecules in the crystal to be a significant factors that needed to be incorporated in the simulations to adequately reproduce the experimental spectra.

\begin{acknowledgments}
We thank G.~Joshi and K.~Jagannathan for useful conversations and advice, A.~Anderson for assistance with use of the computing cluster, J.~Kubasek for assistance with design and machining of the rotation apparatus, and R.~Winn for 3D printing of portions of the apparatus. Support for this work was provided by the U.S.~National Science Foundation under Grant Nos.~DMR-1310135 and DMR-1708692, and by the Amherst College Dean of Faculty. J.R.F.~acknowledges the support of the Amherst College Senior Sabbatical Fellowship Program, funded in part by the H.~Axel Schupf '57 Fund for Intellectual Life. R.A.A.~Cassaro thanks FAPERJ and CNPq for financial support. T.A.~Costa acknowledges CNPq for the fellowship.
\end{acknowledgments}

\bibliographystyle{apsrev4-1}
\bibliography{Rotation_Paper_Bib}

\end{document}

\title{Supplementary Information for Direct Spectroscopic Observation of Berry Phase Interference in the Ni$_4$ Single-Molecule Magnet}

\author{Brendan C.~Sheehan}%$^{1,2}$}%
\affiliation{Department of Physics and Astronomy, Amherst College, Amherst, MA 01002, USA}
\affiliation{Department of Physics, University of Massachusetts Amherst, Amherst, MA 01003, USA}%
\author{Robert Kwark}%$^{2}$}
\affiliation{Department of Physics and Astronomy, Amherst College, Amherst, MA 01002, USA}
\author{Charles A.~Collett}%$^{2}$}
\affiliation{Department of Physics and Astronomy, Amherst College, Amherst, MA 01002, USA}
\affiliation{Department of Physics, Muhlenberg College, Allentown, PA 18104, USA}
\author{Thomaz A. Costa}
\affiliation{Instituto de Qu\'imica, Universidade Federal do Rio de Janeiro, Rio de Janeiro, RJ 21941-909, Brazil}
\author{Rafael A.~All\~{a}o Cassaro}
\affiliation{Instituto de Qu\'imica, Universidade Federal do Rio de Janeiro, Rio de Janeiro, RJ 21941-909, Brazil}
\author{Jonathan R.~Friedman}%$^{1,2}$}
\affiliation{Department of Physics and Astronomy, Amherst College, Amherst, MA 01002, USA}
\affiliation{Department of Physics, University of Massachusetts Amherst, Amherst, MA 01003, USA}

\date{\today}

\maketitle

Here we review some details of data acquisition, analysis and calculations to support the results presented in the main text.

\section{Data Acquisition and Analysis}
\begin{figure}[!htb]
\includegraphics[width=1\textwidth]{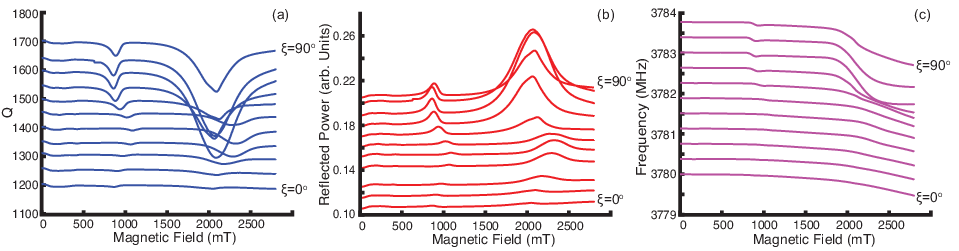}

\caption{Waterfall plots of raw data at 2.0~K: (a) quality factor $Q$, (b) reflected power $P$, (c) resonant frequency $f_{res}$, each as a function of applied magnetic field.  Different curves correspond to different values of the angle $\xi$, with each curve corresponding to a 9$^\circ$ change in $\xi$ from its neighbor.  The vertical axes of each panel gives accurate values for the lowest curve; all others have been shifted up.}\label{QPf_RawData}
\end{figure}

\begin{figure}[!htb]
\includegraphics[width=0.8\textwidth]{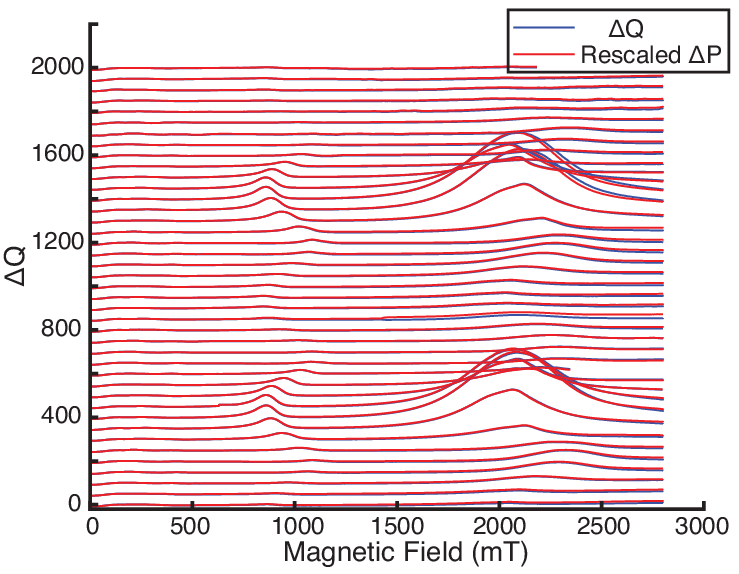}

\caption{Waterfall plot of $\Delta Q$ (blue) and $\Delta P$ (red) as a function of $B$ for data taken at 2.0~K.  Each successive curve corresponds to a 9$^\circ$ increment in the rotation angle $\xi$ from neighboring curves. The vertical axis gives accurate values of $\Delta Q$ for the lowest curve; all others have been shifted up.  The values of $\Delta P$ have been scaled by a common factor to achieve near overlap with the corresponding curve for $\Delta Q$.}\label{PowQWater}
\end{figure}

Data from the sample was acquired using a Keysight E5063A Vector Network Analyzer (VNA) by measuring the reflected power coming from our loop-gap resonator.  The VNA was set to monitor the resonance of the resonator, which produced a Lorentzian lineshape (not shown).  The resonance quality factor ($Q$), reflected power at resonance ($P$) and resonant frequency ($f_{\text{res}}$) were monitored as the experimental variables, notably magnetic field $B$, temperature $T$ and rotation angle $\xi$, were varied. No field modulation was used in the experiments. Figure~\ref{QPf_RawData} shows example spectra of each of the measured resonance variables as a function of $B$.  The data show that the experiment is in the perturbative, linear-response regime in which changes in $Q$ follow changes in $P$.  To confirm this, we plot $\Delta Q= Q_{\text{back}}-Q$ and $\Delta P= P-P_{\text{back}}$ as a function of $B$ on top of each other in Fig.~\ref{PowQWater}, using a scaling factor to match the amplitudes of the signals.  Here $Q_{\text{back}}$ and $P_{\text{back}}$ are the background values (far from any resonant feature) of $Q$ and $P$, respectively.  The fact that the two sets of data are nearly indistinguishable corroborates the linearity of the system's response.  Fitting can be done to either variable.  In the main text of the paper, we choose to use $\Delta P$ as our signal.

\section{States and Transition Matrix Elements}

\begin{figure}[!htb]
\includegraphics[width=1\textwidth]{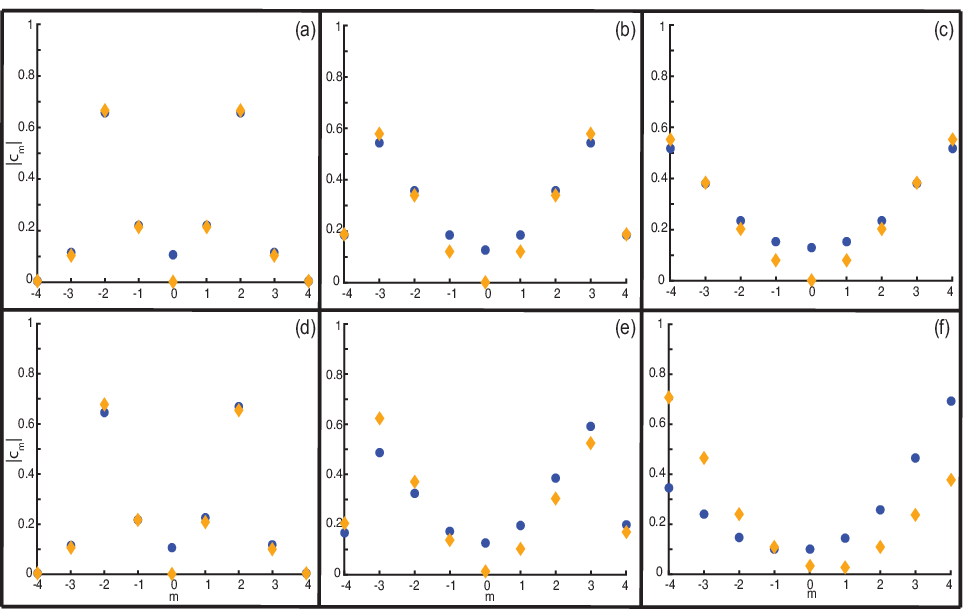}
\caption{Eigenstate decomposition for the transitions observed in the experiments.  Energy eigenstates are represented in the basis of $S_z$, with its  $m=-4,-3,\ldots,4$ eigenvalues.  Here values of $\left|c_m\right|$ are plotted as a function of $m$.  (a), (b) and (c) each show the eigenstates involved in the $m=$2, 3 and 4 transitions, respectively, when the applied field is perpendicular to the z axis.  Blue circles (yellow diamonds) are for (anti)symmetric states.  (d), (e) and (f) show the corresponding states when the field is misaligned by 0.3$^\circ$, i.e.~the sample sees a small z component of field, and the symmetry of the states is lifted.}\label{eigen}
\end{figure}

In understanding our data and performing simulations, we necessarily diagonalize the system's spin Hamiltonian, Eq.~1 (main text) to find the energy eigenstates.  Working in the $S_z$ eigenbasis, each state can be represented in the form $\sum_{m=-4}^4 c_m\ket{m}$.  In Fig.~\ref{eigen}, we show the values of $\left|c_m\right|$ as a function of $m$ for several relevant states, each calculated at the field at which the relevant ESR transition is observed.  Panel (a) shows the two states involved in what we have dubbed the $m=2$ transition because the largest components of of the eigenstates are the $\ket{m=2}$ and $\ket{m=-2}$ states.  The calculation for this panel was done with the field perpendicular to the z axis, resulting in the states being symmetric (blue circles) and antisymmetric (yellow diamonds) in $m$; to wit, $c_m=\pm c_{-m}$.  When the field is tilted by 0.3$^\circ$ so that the there is a small z component of the field (panel (d)), the symmetry of the states is lifted, meaning that the states become more localized in values of $m$.  Panels (b) and (e) show the results of similar calculations for the two states involved in the $m=3$ transition, while panels (c) and (f) correspond to the states involved in the $m=4$ transition, yielding qualitatively similar conclusions with more severe localization produced by misalignment for larger $m$.  

From the eigenstates we calculate the transition matrix elements for the $S_z$ operator, the relevant operator for the experiment since the RF field is applied nearly parallel to the sample's easy axis.  To perform this calculation, we begin by diagonalizing the spin Hamiltonian in the $S_z$ eigenbasis, as described above. As a first approximation, the matrix elements for the $m$th transition can be calculated to be
\begin{equation}
\begin{aligned}
    _m\bra{-}S_z\ket{+}_m &=  \frac{1}{2}\Big(\bra{+m}-\bra{-m}\Big)S_z\Big(\ket{+m}+\ket{-m}\Big) \\
    &=\frac{1}{2}\Big[m\langle+m|+m\rangle + m\langle-m|-m\rangle\Big] \\
    &=m. \\
\end{aligned}
\end{equation}

\begin{figure}[!htb]
\includegraphics[width=1\textwidth]{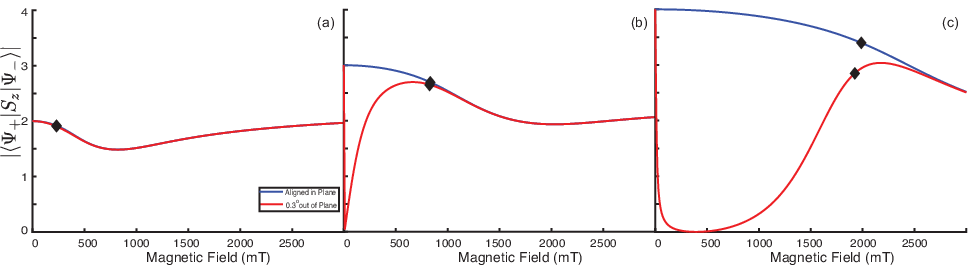}
\caption{Transition matrix elements between relevant states for each observed transition, as a function of magnetic field.  The blue curves correspond to the field being perpendicular to the z axis and the red curves correspond to the field misaligned by 0.3$^\circ$.  The black symbols mark the field at which the resonance condition is fulfilled and the transition is observed.  (a), (b) and (c) give the matrix elements for the 2, 3 and 4 transitions, respectively.}\label{matrix}
\end{figure}

Exact values of the matrix elements can be easily obtained numerically, using the states $\ket{\Psi_+}$ (blue in Fig.~\ref{eigen}) and $\ket{\Psi_-}$ (yellow) found from diagonalizing the Hamiltonian.  Fig.~\ref{matrix} shows the matrix elements for each of the three transitions as a function of DC magnetic field when the field is perpendicular to the easy axis (blue) and misaligned by 0.3$^\circ$ (red).  The symbols indicate the fields at which each transition occurs. Panels (a), (b) and (c) give the matrix elements for the 2, 3 and 4 transitions, respectively.  It is noteworthy that in each case, at zero field the matrix element has a value close to $m$, confirming the approximate calculation above.  With the field perpendicular to z (blue), the matrix element changes smoothly as the field is increased.  In contrast, for the $m=3,4$ transitions with a small misalignment of 0.3$^\circ$ (red), the matrix element drops abruptly for increasing field as the states are localized.  As the field increases further, the tunnel splitting is increased and the states become increasingly less localized.  At the experimental field (markers), the matrix element has again become substantial, leading to a strongly measurable transition.  The difference in matrix element between the red and blue curves qualitatively accounts for the variation in signal amplitude as the angle $\xi$ is varied in the experiment and the sample is rotated from in the (X-Y) plane perpendicular to the field to slightly misaligned and back.

